\documentclass[preprint2]{aastex}

\newcommand{\rvir}{R_{\rm{vir}}}

\newcommand{\msun}{M_{\odot}}

\shorttitle{The Magellanic Stream}
\shortauthors{Besla et al.}

\begin{document}

\title{Simulations of the Magellanic Stream in a First Infall Scenario}
\author{G. Besla\altaffilmark{1}, N. Kallivayalil\altaffilmark{2}, L. Hernquist\altaffilmark{1}, R. P. van der Marel\altaffilmark{3}, T.J. Cox\altaffilmark{4}, D. Kere\v{s}\altaffilmark{1}}
\altaffiltext{1}{Harvard-Smithsonian Center for Astrophysics, 60 Garden Street, Cambridge, MA 02138}
\altaffiltext{2}{MIT Kavli Institute for Astrophysics \& Space Research, 70 Vassar Street, Cambridge, MA 02139}
\altaffiltext{3}{Space Telescope Science Institute, 3700 San Martin Drive, Baltimore, MD 21218}
\altaffiltext{4}{Carnegie Observatories, 813 Santa Barbara Street, Pasadena, CA 91101}
\email{gbesla@cfa.harvard.edu}

\begin{abstract}

Recent high precision proper motions from the Hubble Space Telescope (HST) suggest that the 
Large and Small Magellanic Clouds (LMC and SMC, respectively) 
are either on their first passage or on an eccentric long period ($>$6 Gyr) orbit about the 
Milky Way (MW).  This differs markedly from the canonical picture in which the Clouds travel on a quasi-periodic 
orbit about the MW (period of $\sim$2 Gyr). Without a short period orbit about the MW, the origin of the Magellanic Stream,
a young (1-2 Gyr old) coherent stream of HI gas that trails the Clouds $\sim$150$^{\circ}$ across the sky, can no longer 
be attributed to stripping by MW tides and/or ram pressure stripping by MW halo gas. 
We propose an alternative formation mechanism 
in which material is removed by LMC tides acting on the SMC before the system is accreted by the MW. 
We demonstrate the feasibility and generality of this scenario using an N-body/SPH simulation with 
cosmologically motivated initial conditions constrained by the observations. Under these conditions we 
demonstrate that it is possible to explain the origin of the Magellanic Stream in a first
infall scenario.   This picture is generically applicable to any gas-rich dwarf galaxy pair infalling 
towards a massive host or interacting in isolation.

  \end{abstract}


\section{Introduction}
\label{sec:Intro}

The Magellanic Clouds are the closest known interacting 
pair of galaxies.  Optical and infrared surveys of the system present the Clouds as two distinct 
objects separated in space by a projected distance of $\sim$20 kpc.  The HI distribution, 
however, paints a different picture.  The Clouds are connected by a low metallicity bridge of 
gas, referred to as the Magellanic Bridge, and share a common gaseous envelope ~\citep{P03, B05}.
 The existence of such features suggests that the Clouds are a 
binary interacting pair. The Bridge in particular indicates that they have had a close 
encounter in the recent past.  ~\citet{Toomre} demonstrate that 
 two isolated galaxies are capable of 
removing substantial amounts of material via tides, forming pronounced features such as 
bridges and tails.   Interestingly, the system also possesses a substantial trailing HI 
component, known as the Magellanic Stream ~\citep{M74}. The Stream is a 
filamentary feature of HI gas ~\citep[no stars;][]{G98} that trails behind the Clouds 
for at least 150$^{\circ}$ across the sky ~\citep{B04, N10}.  
The Stream has historically been explained as the product of a tidal and/or hydrodynamic 
interaction between the Clouds and the MW  ~\citep[e.g., GN96, ][]{Co06, M05}.   
This picture stems from the belief that the 
 Clouds have traveled in an orbit that afforded multiple close passages between 
the Clouds and our Galaxy. However, recent HST proper motion (PM) measurements of the 
Clouds by ~\citet{K1,K2} (hereafter K1 and K2), independently confirmed by ~\citet{P08},
 have challenged this picture.  These studies suggest
  instead that the Clouds have, at best, completed one orbit about the MW or 
may even be on their first passage ~\citep[][hereafter B07]{B07}.  This calls for 
a revised interpretation of the origin of the Stream.  

% --> what HST PMs mean

K1 determined a velocity of 378 $\pm$ 18 km/s  for the LMC, which is 80 km/s higher than 
that derived from theoretical models of the Stream (GN96).  B07 showed 
that a backward integration scheme ~\citep[e.g.,][]{M80} using the new velocities 
and an isothermal sphere model for the MW yields an orbit for the LMC with an 
apocenter  $>$200 kpc.  It is, however, unlikely that the rotation curve of the MW remains flat 
out to such distances ~\citep[e.g.,][]{X08}.  If instead a more cosmologically motivated profile
is employed ~\citep[e.g. NFW or Hernquist,][]{NFW,H90} then 
 orbital solutions with multiple pericentric passages are ruled out.  
In particular, if an NFW model is adopted with a mass of $10^{12} \msun$, B07 conclude that 
the Clouds have just experienced their {\it first} close passage past the MW.   

Given the uncertainties in the adopted MW model, it is also possible that the Clouds have 
completed at most one orbit about our Galaxy within a Hubble time. This occurs if either 
the MW's mass were higher ($\sim 2 \times 10^{12} \msun$; B07), 
or if the velocity inferred from the PM measurements were substantially lower, e.g. 
if the velocity at the Solar circle were higher ~\citep{S09,R09}.  In this 
case, the only previous pericentric passage about the Galaxy would be $\sim$6 Gyr ago and 
the apocenter of the orbit would be $\sim$400 kpc (i.e. larger than the virial radius of the 
MW). These values are lower limits since these studies assume the MW's mass is constant 
over time, whereas in the current LCDM paradigm our Galaxy is believed to have been half as 
massive $\sim$8 Gyr ago ~\citep{W02}. Depending on the mass evolution of the MW it may be impossible
for the Clouds to have completed multiple passages.

Independent of which of the two orbital scenarios outlined above is correct, there is no orbital solution that 
brings the Clouds near the MW over the past 3 Gyr.  However, there is strong 
evidence that the Stream is a young feature (1-2 Gyr). Estimates of the survivability of high velocity clouds by ~\citet{H09, Keres} 
make it improbable for the Stream to have survived much longer.  The Stream also exhibits surprisingly high H$\alpha$ 
emission ~\citep[ $\sim$750 mR;][]{W96}, which implies that cloudlets within the Stream are being ablated away on a 100-200 Myr 
timescale and must be continuously replenished ~\citep{BH07}.  
The lifetime of the Stream poses a problem for all past numerical models, which invoke some combination of
MW tides and/or ram pressure stripping to form the Stream. These models require at least one complete orbit about the MW, implying an incompatible age of at least 6 Gyr. 

As such, regardless of 
whether the Clouds are on their first or second passage about the MW, in the context of the 
origin of the Stream we are left with the same problem:  How can the Stream have 
formed without a complete orbit about the MW? 
Or, more generically, how can a pronounced tail be formed from a pair of dwarfs on their first infall towards a massive host?

%-- WHAT I'M GOING TO ASSUME

Based on the above considerations, we explore the following scenario using simulations. 
We assume that the Clouds were a stable binary system 
and adopt the simplifying hypothesis that the Clouds are on their first passage about the MW 
in order to illustrate how the Stream can form
% interactions between the Clouds themselves 
%could have produced the Stream 
without relying on a close encounter with the MW.  
 We postulate that the Magellanic Stream and Bridge are in fact a classical bridge and tail caused by the tidal interaction 
between the Clouds before they have been accreted by the 
MW.  The MW potential shapes the orbit of the Clouds and thereby controls the appearance 
of the tail, causing the line-of-sight velocities and spatial location of the tail to be as observed 
in the Stream today.

Given these assumptions we attempt to explain the following observed features of the Stream:  
1) The absence of stars; 2) The $\sim$150$^{\circ}$ extent; 3) The 
spatial location projected on the plane of the sky; 4) The line-of-sight velocities along
its length; 5) The HI column densities; 
and 6) The pronounced asymmetry between the trailing and leading components.

\section{Methodology}
\label{sec:Method}

Simulations were carried out using the N-body/SPH code
GADGET2 (Springel et al 2005).  Star formation is not included for simplicity, but the gas is allowed to cool radiatively.

We model the MW as a static NFW potential of mass of $1.5  \times 10^{12} \msun$, $\rvir$ =240 kpc
 and a concentration parameter of 12. Dynamical friction from the MW halo is not explicitly accounted
 for, but is expected to have little impact on the orbit in a first passage (see B07, figure 4).
The L/SMC are both modeled using Hernquist profiles 
for their dark matter content, and exponential gaseous and stellar disks, where the scale length 
of the gas disk is 6 (LMC) or 5 (SMC) times that of the stellar disk. 
Such extended gaseous disks are not atypical for isolated dwarf galaxies ~\citep{S02}. %,Beg05}.

The stellar and gas masses for the Clouds are well constrained within their respective
 observable limits. However, the total dark matter content of these galaxies is unknown. 
  All previous 
  models of the Magellanic system have assumed the LMC is tidally truncated to a radius of 
  15 kpc ~\citep{vdM02}, resulting in a total mass estimate of 2-3 $\times 10^{10} \msun 
  $  ~\citep[e.g., GN96, ][]{M80, Be05}.  On an orbit 
  where MW tides are largely inconsequential, the L/SMC will not be truncated. 
 Instead, we use current halo occupation models to relate the observed stellar mass of the LMC to its original
halo mass before infall into the MW halo ~\citep{B03,C06,W06}.
In Table ~\ref{Table:ICs}  we summarize our adopted model properties. 
The L/SMC are found to have infall masses an order of magnitude larger than employed 
in previous models.

\section{Simulation Results and Comparisons with Data}
%\section{Binary L/SMC orbit}
\label{sec:LS}

We believe that the key to understanding the origin of the Stream lies in 
understanding the interaction history of the Clouds themselves.
The majority of previous models of the Stream have assumed that the SMC is in a circular 
orbit about the LMC (separation of $\sim$20 kpc).  Using the cosmologically expected infall 
masses, the dynamical friction timescale for such an orbit would be much less than a Hubble 
time, so the SMC's orbit about the LMC cannot be circular. The idea that collisions between 
the Clouds have caused material to be loosely bound to the Clouds is not novel ~\citep[GN96, ][]{H94}.
 However, the morphology resulting from 
collisions between the Clouds as the SMC travels in a {\it highly eccentric} orbit about the LMC, 
 independent of the MW,  has never been explored.

\begin{figure*}[htb]
\begin{center}
\resizebox{0.8\hsize}{!}{\includegraphics{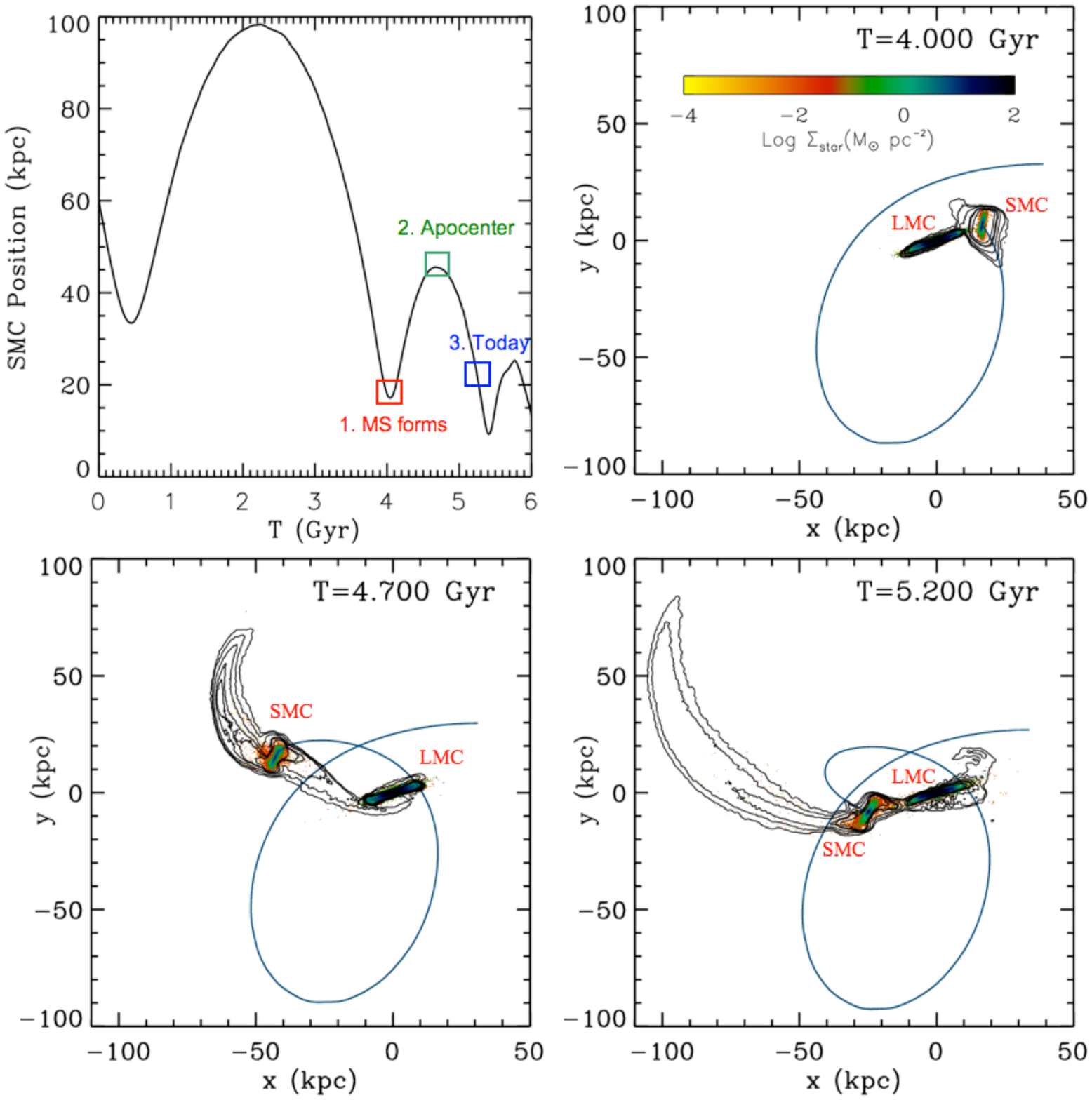}}
\end{center}
\caption{The orbit of the SMC about the LMC (without the MW).   Top left: The separation 
between the Clouds as a function of time.
  The chosen orbit is highly eccentric (e=0.7) and decays rapidly owing to dynamical friction.  
  The Stream forms at 4 Gyr (red box) and the time today would correspond to $\sim$5.2 Gyr (blue box). 
   Subsequent Panels: The HI gas column density is plotted as contours over the stellar distribution for 
   specific points in the orbit.
   % during the SMC's orbit about the LMC. 
   Gas contours span a range of 
   $10^{18}-10^{20}$ /cm$^2$, where each contour represents an increase in column density by a factor 
   of 1.5. HI gas is identified as gas at temperatures below 12000 K and column densities larger than
    $10^{18}$ /cm$^2$, although a background ionizing field is not included. 
    A bridge of gas connecting 
    the Clouds and a 100 kpc long gaseous stream with no stellar counterpart is formed without the aid 
    of MW tides or ram pressure.  }
\label{fig:LS}
\end{figure*}

We have used the current projected separation between the Clouds (~23 kpc), and the 
observed relative velocity ($\sim105 \pm 42$ km/s; K2) to constrain the orbital history of the 
SMC about the LMC.  
 The resulting orbit is nearly parabolic (e= 0.7), with an 
apocenter of $\sim100$ kpc  (see Figure ~\ref{fig:LS}). The SMC disk is oriented 90$^{\circ}$ with respect 
to its orbital plane about the LMC. 
If the SMC were in a coplanar, retrograde orbit about the LMC, no material would be removed with this orbital
configuration. This implies that dispersion supported material (e.g. dark matter or stellar halo) will be unaffected within 
the SMC's disk radius.

%If the SMC's disk were coplanar and retrograde with
%respect to its orbit about the LMC, 
%{\it no} material would be removed from the SMC's disk with this orbital configuration
% (r$_{peri} \sim 20$ kpc). 
% This implies that any dispersion supported material (e.g. dark matter or stellar halo components)
% will be unaffected by LMC tides within the disk radius. 

 The presented orbital solution is not unique;
however, a highly eccentric orbit is required to prevent the Clouds from merging. Moreover, these orbits
are cosmologically typical ~\citep[see, e.g., ][]{Benson,Wetzel}.  Any eccentric orbit that allows for   % KB06
high speed encounters between the Clouds will yield similar bridge and tail structures. 

We simulated the interaction between the Clouds as the SMC travels along the orbit 
shown in Figure 1 (top left) starting $>$ 6 Gyr ago. At each close passage between the 
Clouds, gas is removed from the SMC by LMC tides, forming a tidal tail and bridge. However, 
as the Clouds move apart, the tidal material falls back towards the L/SMC disks. 
These features are thus transient phenomena. 
Little material is removed from the LMC, despite its
extended gaseous disk component:  SMC tides are ineffective. 

The pericentric passage at 4 Gyr in Figure 
~\ref{fig:LS} (top left) results in significant gas removal from the outer regions of the SMC's 
extended gas disk, whereas the smaller stellar disk remains intact.
As the SMC travels away from the LMC, material that was tidally removed by the LMC stretches
our to distances as large as 100 kpc 
from the Clouds. This material will eventually form the Stream (Figure ~\ref{fig:LS}, 
bottom right).  Hence, an extended gaseous tail 
is produced without the aid of MW tides or ram pressure.

We assume that the Clouds have been an interacting pair for a 
significant fraction of a Hubble time and have crossed the virial radius (240 kpc) of our 
Galaxy $\sim$1 Gyr ago.  From Figure ~\ref{fig:LS}, 
 a 100 kpc long tail is formed after $\sim$5 Gyr. We therefore
 stop the simulation 1 Gyr earlier (i.e. at the timestep 
 corresponding to the bottom left panel of Figure ~\ref{fig:LS}) and place the 
 binary system outside the virial radius of the MW.  We then allow the Clouds to travel to their 
 current observed locations on an orbit consistent within 1$\sigma$ of the PMs of 
 the LMC  ~\citep[K1,][]{P08}.  We did not attempt to reproduce the SMC PM 
 determined by K2 and ~\cite{P08} since they are discrepant. However, the resulting
  SMC line-of-sight velocity agrees well with the observed value.

\begin{figure*}[htb]
\begin{center}
\resizebox{0.8\hsize}{!}{\includegraphics{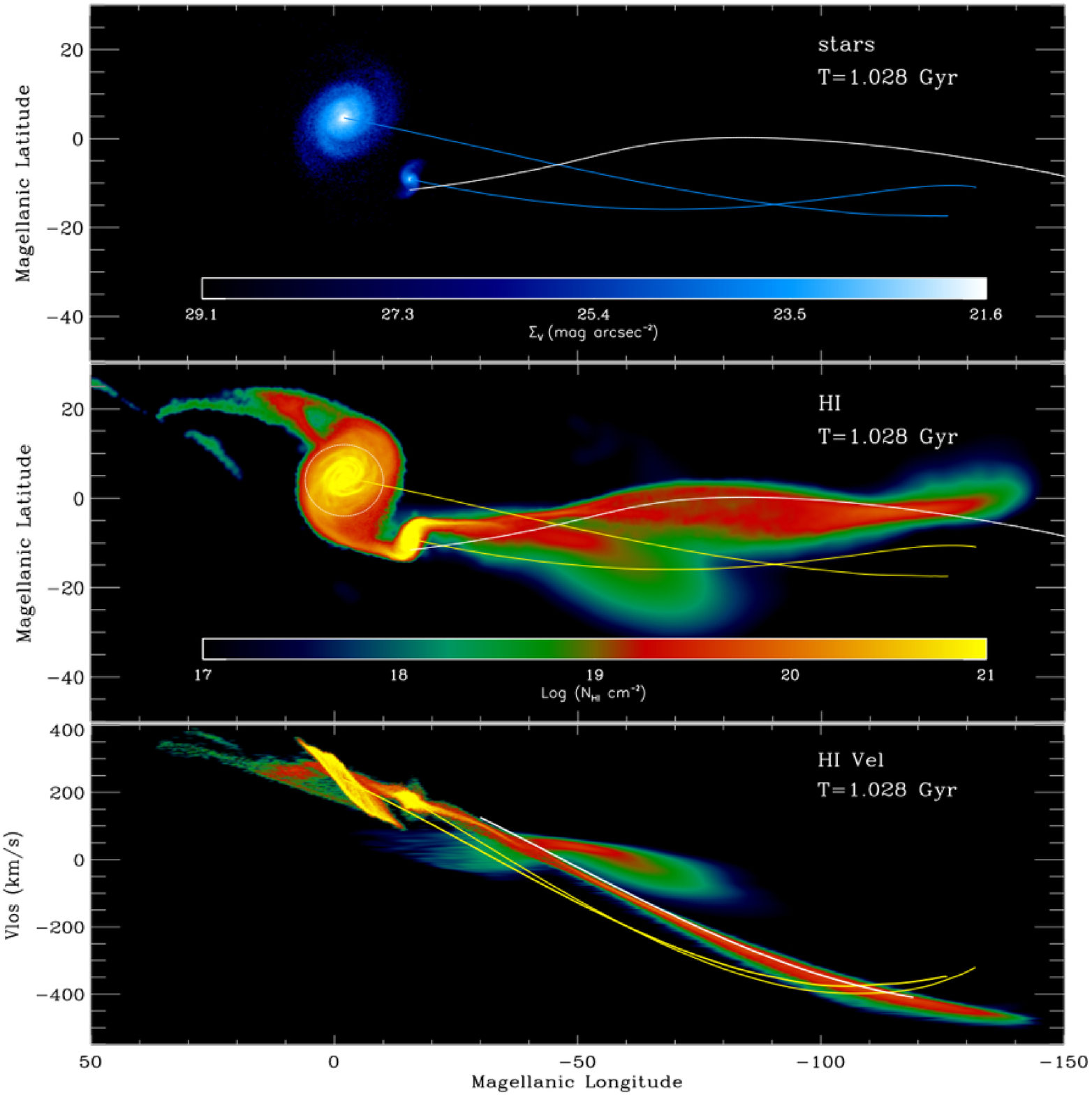}}
\end{center}
\caption{The stellar surface brightness, HI gas column densities and line-of-sight velocities of the simulated 
Magellanic system. Top panel: The resulting stellar distribution is projected in Magellanic coordinates
(N08), a variation of the Galactic coordinate system where the Stream is straight. The distribution is 
color-coded in terms of Vband surface brightness. % (Kband mag/arcsec$^2$; the 2MASS limit is 20 mag/arcsec$^2$). 
 The past orbit of the L/SMC are indicated by the blue lines. 
Middle panel: The HI gas column densities of the simulated stream range from 
$10^{18}-10^{21}$ /cm$^2$, as expected ~\citep{P03}. 
The white circle indicates the observed extent of the LMC's HI disk: the simulated LMC is more extended than observed,
indicating ram pressure likely plays a role to truncate the disk.
In both the top and middle panels, the solid white line indicates the past orbit of the SMC according 
to the old theoretically derived PMs (GN96) which was a priori chosen to trace the Stream on the plane of the sky.
 The true orbits (determined by all PM
measurements) for the L/SMC are indicated by the yellow lines. 
  Bottom panel: The line-of-sight velocities along the simulated stream are 
  plotted and color-coded based on HI column 
density, as in the middle panel. The white line is a fit to the observed data (N08). The LMC disk is too extended, causing a larger 
velocity spread than observed.
The line-of-sight velocities along the past orbits of the L/SMC are indicated by the 
yellow lines, which do not follow the true velocities along the Stream (e.g. B07 figure 20). 
 The Stream is kinematically distinct from the orbits of the Clouds.}
 \label{fig:MS}
\end{figure*}

In Figure ~\ref{fig:MS} we show the resulting stellar distribution (top) and HI gas column density map (middle) for 
our simulated stream: there are no observable stars in the simulated stream and the stream extends $\sim$150$^{\circ}$.  
The white 
line indicates the current location of the observed Stream and is well matched by the 
simulation. Notice that the past orbits (yellow lines) are not co-located with the simulated 
stream, as expected since the north component of the LMC PM vector is not 
aligned with the Stream (see B07, figure 9).  This spatial mismatch is a natural result of our 
model: material is removed from the SMC along the LMC-SMC binary orbital plane, which is 
not co-planar with the orbital plane of the Clouds about the MW. 
 The SMC disk was oriented $90^{\circ}$ to the L/SMC binary plane in order to maximize this offset. 
 This further implies that the SMC is seen $\sim$edge on from our viewing perspective, explaining its
 surprisingly large observed line-of-sight depth ~\citep[$\sim$5 kpc; ][]{SS09}.  %The simulated SMC spans a line of sight depth of $\sim$5-10 kpc for STARS. 

Figure ~\ref{fig:MS} (bottom) illustrates the resulting line-of-sight velocities for the simulated 
stream. The white line is a fit to the data  ~\citep{P03, N08}[hereafter, N08] and shows good agreement with the simulation. 
The line-of-sight velocities along 
the new orbits (yellow lines) are much larger than those observed along the 
Stream: since the Stream and the orbits are not co-located, their line-of-sight velocities are 
not similar (see also B07, figure 20).  The MW's gravitational field serves to stretch the stream and 
modify its galactocentric distance: 
the tip of the simulated stream is located $\sim$100-140 kpc away.  % away from our Galaxy. 
Using these distance estimates, we find that the mass in our simulated stream 
matches the observed value \citep[$4.8 \times 10^{8} \msun$;][]{B05} to within a factor of 2. 

The simulated gas column densities range from $10^{18}-10^{21}$ /cm$^2$ as observed,  
although the exact column density gradient along the length of the Stream is not reproduced.  
The column density is not homogeneous across the width of the simulated stream: the 
inclusion of metal cooling, ionization and confinement/interaction by/with the ambient MW halo gas will likely 
aid in reproducing the bifurcated, filamentary nature of the observed Stream.  

The leading component of the simulated stream is much smaller than the trailing component, as 
observed.  This arises because the leading tidal arm from the SMC falls towards the 
LMC, while the tail continues to grow (Figure ~\ref{fig:LS}).  However, the leading component in the model
 is not in the correct location on the sky when compared to observations. 
 This material leads the orbit, which is not aligned with the Stream. Consequently, this apparent problem 
will also occur in the traditional tidal models of e.g. ~\citet{Ru09, Be08, Co06, Be05} and  ~\citet[][hereafter, GN96]{GN96}.  
This indicates that hydrodynamic processes such as ram 
pressure ~\citep{M05} are needed to shape the final appearance of the simulated stream.

\section{Discussion and Conclusions} 
\label{sec:Conc}
We have shown that tidal interactions between the Clouds are sufficient to 
remove a substantial amount of material from the SMC without the aid of MW tides or hydrodynamic 
interactions. Our goal was not to reproduce every detail of the Stream, but merely to show that a plausible Stream model 
can be formed without a previous passage about the MW. Nevertheless, many features of the simulation fit
the data remarkably well. 

We have explained the absence of an observable (brighter than Vband 27 mag/arcsec$^2$) stellar counterpart to the
 Stream by placing the SMC
 on an orbit about the LMC with a large impact parameter, which allows its compact stellar disk to remain intact ~\citep[see also, ][]{YN03}. 
The 150$^{\circ}$ extent of the Stream is reproduced owing to the eccentric orbit of the SMC about the LMC (apocenter $\sim$100-50 kpc)
 and MW tides after the system has been accreted. Such 
an eccentric orbit is likely only viable in a first infall scenario, as MW tides would have 
disrupted the system at the previous pericentric approach.  
The spatial location of and velocity gradient along the simulated stream are correct and offset from those of the orbits: 
  material is removed in the L/SMC binary plane, which is not coplanar with the orbit of the Clouds about the MW. 
 The simulated HI column densities also match the observations, although the exact 
 column density gradient is not reproduced without additional physics.  Finally, the strong asymmetry
 between the leading and trailing stream components is a natural consequence of a classical tidal bridge and tail scenario.

There are, however, other features that require additional physics to address, such as the claim of correlated
bursts of star formation in both Clouds ~\citep{Harris09}. In our new scenario, correlated bursts of star formation may correspond
to close passages between the Clouds rather than with the MW. Furthermore, without ram 
pressure and stellar feedback we cannot explore the claim by N08 that 
half of the Stream originates from the LMC as a stellar outflow ~\citep[see also, B07 and ][]{Olano}.  
 We intend to examine these issues and more detailed comparisons of our simulations to the Magellanic system in future studies. 
 However, we note that recent metallicity measurements at the tip of the Stream ~\citep{F10} indicate that 
the Stream is extremely metal poor and thus inconsistent with a stellar outflow scenario. \citet{F10} also find that the
 oxygen abundance at the tip is more consistent with that of 
the SMC rather than the LMC. As such, a model where the Stream originates primarily from the SMC is not ruled out by observations of the Stream. 

The simulation results presented here with respect to the L/SMC interaction have broader applicability 
than just to the Magellanic system. 
Within the current LCDM paradigm halos at all scales are expected to build up their mass 
hierarchically.  Interacting dwarf galaxies are therefore 
cosmologically expected both in isolation and within larger halos ~\citep{S10, D09, K06}. The isolated interacting 
 Magellanic type galaxies NGC 4490/85, which are surrounded by an extended HI envelope ~\citep{Clemens}, and the interacting
 M51/NGC 5195 pair ~\citep{M51} may be observational analogs of our initial L/SMC system.
% may be an observational analog of our isolated L/SMC initial conditions.  M51 and NGC 5195 
 %may also be a more massive analog of this type of interaction ~\citep{M51}. 
The presented model thus illustrates that dwarf-dwarf galaxy tidal interactions are a powerful mechanism to 
morphologically change dwarf galaxies without the need for repeated interactions with a massive host.

\clearpage

\begin{deluxetable}{ccc}
 \tablecolumns{3}
 \tablewidth{0pc}
 \tablecaption{L/SMC Initial Conditions Adopted}
 \tablehead{
 \colhead{Property} &  \colhead{LMC} & \colhead{SMC} \\  } %& \colhead{Ref.} \\ }
 \startdata
 M$_\ast$ ($\msun$) &  $2.2 \times 10^{9}$ & $1.3\times 10^8$ \\  %  & \citet{vdM02}\\
 M$_{gas}$ ($\msun$)  & $1.4 \times 10^{9} $ & $1.1\times 10^9$ \\ %  & \citet{B05} \\
% &   	  M$_{halo}$\tablenotemark{a} & 150-200 & \citet{W06}\\
 M$_{halo}$ ($\msun$) \tablenotemark{a}& $1.8\times 10^{11} \msun$ & $2.5 \times 10^{10}$ \\ %& Adopted in this work \\
 V200 (km/s)  &  82 & 42 \\
 Concentration &  9 & 15 \\
 Stellar scale length (kpc) &  1.7 & 0.7 \\
 Gas scale length (kpc) &  10.2 & 3.6  \\   
 Nstars  & 100000  & 100000\\
 Ngas & 300000 & 300000\\
 Nhalo & 100000 & 100000 \\
% SMC &  M$_\ast$ &  $1.3\times 10^{8} \msun$ \\ % & \citet{S04} \\
% &  	  M$_{gas}$ & $1.1\times10^{9} \msun $ \\  % 0.4  &  Adopted in this work \citet{B05} \\
 %&      M$_{stream}$ & 0.48 & \citet{B05} \\
 %&  	 M$_{halo}$\tablenotemark{a} & 20-50 & \citet{W06}, no stream \\
% &      &    20-80\tablenotemark{b}  & \citet{W06}, including stream \\
% &    M$_{halo}$ \tablenotemark{a} & $2.5 \times 10^{10}$ \msun $ \\ %Adopted in this work \\
% & V200 & 42 km/s \\
% &  Concentration & 15 \\
% & Stellar scale length & 0.7 kpc \\
% & Gas scale length & 3.6 kpc \\
% & Nstars & 100000\\
% & Ngas & 300000 \\
% & Nhalo & 100000 \\
 \enddata
\tablenotetext{a}{The total halo mass is determined using the observed stellar mass of the LMC(SMC) 
M$_\ast$ =  $3 \times 10^{9} \msun$ ($3 \times 10^{8}\msun$) \citep{vdM02, S04} and the relations from ~\citet{W06}.  Note that the observed M$_\ast$ is less than that quoted in this table 
in order to account for star formation in future studies.} 
%The total stellar mass of the LMC(SMC) is less than the observed). 
%This is compensated for in the gas component and engineered to account for star formation over the 6 Gyr timescale
%of the simulation. Star formation isn't turned on in these simulations but will be in the future, allowing for the same ICs to be used in future studies.  
%For the purpose of this analysis it is sufficient that the total baryon mass of the LMC(SMC) is correct within the observable extent.} 
%\tablenotetext{b}{Total halo mass is determined using M$_\ast$ and the relations from Wang et al 2006}
%\tablenotetext{b}{Here the Stream mass is included in the total baryon mass of the SMC}
 \label{Table:ICs}
 \end{deluxetable}


\begin{thebibliography}{}

%\bibitem[Begum et al. (2005)]{Beg05} Begum, A., Chengalur, J.N. \& Karachentsev, I.D. 2005, \aap, 433, L1

\bibitem[Bekki (2008)]{Be08} Bekki, K., 2008, \apjl, 684, L87


\bibitem[Bekki \& Chiba (2005)]{Be05} Bekki , K. \& Chiba, M., 2005, \mnras, 356, 680

\bibitem[Benson (2005)]{Benson} Benson, A. J., 2005, \mnras, 358, 551

\bibitem[Besla et al. (2007)]{B07} Besla, G., Kallivayalil, N., Hernquist, L., Robertson, B., Cox, T.J., van der Marel, R.P., \& Alcock, C., 2007, \apj, 668, 949, [B07]

\bibitem[Bland-Hawthorn et al. (2007)]{BH07} Bland-Hawthorn, J., Sutherland, R., Agertz, O. \& Moore, B., 2007, \apj, 670, L109

\bibitem[Braun \& Thilker (2004)]{B04} Braun, R. \& Thilker, D.A., 2004, \aap, 417, 421

\bibitem[Br\"{u}ns et al. (2005)]{B05} Br\"{u}ns, C., Kerp, J., Staveley-Smith, L., Mebold, U., Putman, M.E., Haynes, R.F., Kalberla, P.M., Muller, E. \& Filipovic, M.D., 2005, \apj, 432, 45

\bibitem[Clemens et al. (1998)]{Clemens} Clemens, M.S., Alexander, P. \& Green, D.A., 1998, \mnras, 297, 1015

\bibitem[Connors et al. (2006)]{Co06} Connors, T.W., Kawata, D. \& Gibson, B.K., 2006, \mnras, 371, 108

\bibitem[Conroy et al. (2006)]{C06} Conroy, C., Wechsler, R.H. \& Kravtsov, A.V., 2006, \apj, 647, 201

\bibitem[D'Onghia et al. (2009)]{D09} D'Onghia, E., Besla, G., Cox, T.J. \& Hernquist, L., 2009, Nature, 460, 605

%\bibitem[D'Onghia \& Lake (2009)]{D08} D'Onghia, E. \& Lake, G., 2008, \apj, 686, L61

\bibitem[Fox et al. (2010)]{F10} Fox, A.J., Wakker, B.P., Smoker, J.V., Richter, P., Savage, B.D., \& Sembach, K.R. 2010, arXiv:1006.0974

\bibitem[Gardiner \& Noguchi (1996)]{GN96}  Gardiner, L.T. \& Noguchi, M., 1996, \mnras,  278, 191 [GN96]

\bibitem[Guhathakurta \& Reitzel (1998)]{G98} Guhathakurta, P. \& Reitzel, D.B, 1998, Galalctic Halos: A UC Santa Cruz Workshop, ed. Zaritsky, D., ASP Conf Series, 136, 22

\bibitem[Harris \& Zaritsky (2009)]{Harris09} Harris, J. \& Zaritsky, D., 2009, \aj, 138, 1243

\bibitem[Heller \& Rohlfs (1994)]{H94} Heller, P. \& Rohlfs, K., 1994, \aap, 291, 743

\bibitem[Heitsch \& Putman (2009)]{H09} Heitsch, F. \& Putman, M.E., 2009, \apj, 698, 1485

\bibitem[Hernquist (1990a)]{H90} Hernquist, L. , 1990a, \apj, 356, 359

\bibitem[Hernquist (1990b)]{M51} Hernquist, L., 1990b, Dynamics and Interactions of Galaxies, ed Wielen, R., 108 

\bibitem[Kallivayalil et al. (2006a)]{K1} Kallivayalil, N., van der Marel, R. P., Alcock, C., Axelrod, T., Cook, K. H., Drake, A. J. \& Geha, M., 2006a, \apj,  638, 772 , [K1]

\bibitem[Kallivayalil et al. (2006b)]{K2} Kallivayalil, N., van der Marel, R. P. \& Alcock, C., 2006b, \apj, 652, 1213, [K2]

\bibitem[Kere\v{s} \& Hernquist  (2009)]{Keres} Kere\v{s}, D. \& Hernquist, L., 2009, \apjl, 700, L1

%\bibitem[Khochfar \& Burkert (2006)]{KB06} Khochfar, S. \& Burkert, A., 2006, \aap, 445, 403

\bibitem[Knebe et al. (2006)]{K06}  Knebe, A., Power, C., Gill, S.P.D. \& Gibson, B.K.. 2006, \mnras, 368, 741

\bibitem[Mathewson et al. (1974)]{M74} Mathewson, D.S., Cleary, M.N., \& Murray, J.D., 1974, \apj, 190, 291

\bibitem[Mastropietro et al. (2005)]{M05} Mastropietro, C., Moore, B., Mayer, L., Wadsley, J. \& Stadel, J. , 2005, \mnras, 363, 509

\bibitem[Murai \& Fujimoto (1980)]{M80} Murai, T. \& Fujimoto, M. , 1980, "Astron. Soc. Japan", 32, 581

\bibitem[Navarro et al. (1996)]{NFW} Navarro, J.F., Frenk, C.S. \& White, S.D.M., 1996, \apj, 490, 493

\bibitem[Nidever et al. (2008)]{N08} Nidever, D.L., Majewski, S.R. \& Burton, W.B., 2008, \apj, 679, 432, [N08]

\bibitem[Nidever et al. (2010)]{N10} Nidever, D.L., Majewski, S.R. \& Burton, W.B., 2010, submitted

\bibitem[Olano (2004)]{Olano} Olano, C.A., 2004, \aap, 423, 895

\bibitem[Piatek et al. (2008)]{P08} Piatek, S., Pryor, C. \& Olszewski, E.W. 2008, \apj, 135, 1024

\bibitem[Putman et al. (2003)]{P03} Putman, M. Staveley-Smith, L., Freeman, K.C., Gibson, B.K. \& Barnes, D.G., 2003, \apj, 586, 170

\bibitem[Reid et al. (2009)]{R09} Reid, M.J., Menten, K.M., Zheng, X.W., Brunthaler, A., Moscadelli, L., Xu, Y., Zhang, B., Sato, M., Honma, M., Hirota, T., Hachisuka, K., Choi, Y.K., Moellenbrock, G.A. \& Bartkiewicz, A., 2009, \apj, 700, 137

\bibitem[R\r{u}\v{z}i\v{c}ka et al. (2009)]{Ru09} R\r{u}\v{z}i\v{c}ka, A., Theis, C. \& Palou\v{s}, J.,  2009, \apj, 691, 1807

\bibitem[Shattow \& Loeb (2009)]{S09} Shattow, G. \& Loeb, A., 2009, \mnras, 392, L21

\bibitem[Simha et al. (2010)]{S10} Simha, V., Weinberg, D.H., Dav\'{e}, R., Gnedin, O.Y., Katz, N. \& Kere\v{s}, D., 2010, \mnras, 399, 650

%\bibitem[Smith et al (2010)]{Smith10} Smith, R., Davies, J.I. \& Nelson, A.H. 2010, arXiv:1004.4602

\bibitem[Springel (2005)]{S05} Springel, V., 2005, \mnras, 364, 1105

\bibitem[Stanimirovi\'{c} et al. (2004)]{S04} Stanimirovi\'{c}, S., Staveley-Smith, L. \& Jones, P.A., 2004, \apj, 604, 176

\bibitem[Stanimirovi\'{c} et al. (2008)]{S08} Stanimirovi{\'c}, S., Hoffman, S., Heiles, C., Douglas, K.A., Putman, M.E., \& Peek, J.E.G., 2008, \apj, 680, 276

\bibitem[Subramanian \& Subramanian (2009)]{SS09} Subramanian, S. \& Subramanian, A., 2009, \aap, 496, 399

\bibitem[Swaters et al. (2002)]{S02} Swaters, R.A., van Albada, T.S., van der Hulst, J.M. \& Sancisi, R., 2002, \apj, 390, 829

\bibitem[Toomre \& Toomre (1972)]{Toomre} Toomre, A. \& Toomre, J. 1972, \apj, 178, 623

\bibitem[van der Marel et al. (2002)]{vdM02} van der Marel, R.P., Alves, D.R., Hardy, E. \& Suntzeff, N.B., 2002, \aj, 124, 2639

\bibitem[van den Bosch et al. (2003)]{B03} van den Bosch, F.C., Mo, H.J. \& Yang, X., 2003, \mnras, 345, 923

\bibitem[Wang et al. (2006)]{W06} Wang, L., Li, C., Kauffmann, G. \& De Lucia, G., 2006, \mnras, 371, 537

\bibitem[Wechsler et al. (2002)]{W02} Wechsler, R.H., Bullock, J.S., Primack, J.R., Kravtsov, A.V. \& Dekel, A., 2002, \apj, 568, 52


%\bibitem[van der Marel et al. (2009)]{vdm09} van der Marel, R.P., Kallivayalil, N. \& Besla, G., 2009, The Magellanic system: stars, gas and galaxies, Proceeding of the International Astronomical Union, IAU Symposium, 256, 81

\bibitem[Weiner \& Williams (1996)]{W96} Weiner, B.J. \& Williams, T.B., 1996 \aj, 111, 1156

\bibitem[Wetzel (2010)]{Wetzel} Wetzel, A., 2010, arXiv:1001.4792


\bibitem[Xue et al. (2008)]{X08} Xue, X. X., Rix, H. W.  Zhao, G.,  Re Fiorentin, P., Naab, T.,  Steinmetz, M.,  van den Bosch, F. C.,  Beers, T. C., Lee, Y. S., Bell, E. F., Rockosi, C., Yanny, B.,  Newberg, H., Wilhelm, R.,  Kang, X.,  Smith, M. C. \& Schneider, D. P., 2008, \apj, 684, 1143

\bibitem[Yoshizawa \& Noguchi (2003)]{YN03} Yoshizawa, A. M. \& Noguchi, M., 2003, \mnras, 339, 113




\end{thebibliography}
\end{document}